# Author name disambiguation of bibliometric data: A comparison of several unsupervised approaches

Alexander Tekles[1] and Lutz Bornmann[2]

[1] *alexander.tekles.extern@gv.mpg.de*
Division for Science and Innovation Studies, Administrative Headquarters of the Max Planck Society, Hofgartenstr. 8, 80539 Munich (Germany)
Ludwig-Maximilians-University Munich, Department of Sociology, Konradstr. 6, 80801 Munich, Germany

[2] *bornmann@gv.mpg.de*
Division for Science and Innovation Studies, Administrative Headquarters of the Max Planck Society, Hofgartenstr. 8, 80539 Munich (Germany)

**Abstract**
Adequately disambiguating author names in bibliometric databases is a precondition for conducting reliable analyses at the author level. In the case of bibliometric studies that include many researchers, it is not possible to disambiguate each single researcher manually. Several approaches have been proposed for author name disambiguation but there has not yet been a comparison of them under controlled conditions. In this study, we compare a set of unsupervised disambiguation approaches. Unsupervised approaches specify a model to assess the similarity of author mentions a priori instead of training a model with labelled data. In order to evaluate the approaches, we applied them to a set of author mentions annotated with a ResearcherID, this being an author identifier maintained by the researchers themselves. Apart from comparing the overall performance, we take a more detailed look at the role of the parametrization of the approaches and analyse the dependence of the results on the complexity of the disambiguation task. It could be shown that all of the evaluated approaches produce better results than those that can be obtained by using only author names. In the context of this study, the approach proposed by Caron and van Eck (2014) produced the best results.

**Introduction**

Bibliometric analyses at an individual level depend on the adequate identification of the authors' oeuvres. At best, all of an author's papers should be considered without fail, while other papers should not be falsely assigned to that author. Getting as close as possible to this ideal situation is especially important since poorly disambiguated data may distort the results of analyses at an author level (Kim and Diesner 2016; Kim 2019). Some identifiers that uniquely represent authors are available in bibliometric databases. These, however, are maintained by the researchers themselves (e.g. ResearcherID, ORCID) – implying a low coverage and the possibility of deliberate false assignments – or are based on an undisclosed automatic assignment (e.g. Scopus Author ID) – which does not allow an assessment of the quality of the algorithm (the algorithm is not publicly available). Automatic approaches that try to solve the task of disambiguating author names have thus been proposed in bibliometrics. This task presents a non-trivial challenge since different authors may have the same name (homonyms) and one author may publish under different names (synonyms).

Table 1. Examples for homonyms and synonyms in bibliometric databases

| Publication title | Author name | Author ID |
|---|---|---|
| Social theory and social structure | R. Merton | 1 |
| The Matthew effect in science | Robert Merton | 1 |
| Allocating Shareholder Capital to Pension Plans | Robert Merton | 2 |

Table 1 shows the titles, the author names and an author identifier for three publications, including both homonyms and synonyms. The author names of the first two publications are synonyms since they refer to the same person but differ in terms of the name. The author names of the last two publications are an example of homonyms since they refer to different persons but share the same name.

In this study, we compare four unsupervised disambiguation approaches. In order to evaluate the approaches, we applied them to a set of author mentions annotated with a ResearcherID, this being an author identifier maintained by the researchers themselves. Apart from comparing the overall performance, we take a more detailed look at the role of the parametrization of the approaches and analyse the dependence of the results on the complexity of the disambiguation task.

**Related work**

In order to find sets of publications corresponding to real-world authors, approaches for disambiguating author names try to assess the similarity between author mentions by exploiting metadata such as co-authors, subject category, journal, etc. In order to reduce runtime complexity and exclude a high number of obvious false links between author mentions, most approaches reduce the search space by blocking the data in a first step (On et al. 2005). The idea is to generate disjunctive blocks so that author mentions in different blocks are very likely to refer to different identities, and therefore the comparisons can be limited to pairs of author mentions within the same block (Newcombe 1967; Levin et al. 2012). A widely used blocking strategy for disambiguating author names in bibliometric databases is to group together all author mentions with an identical canonical representation of the author name, consisting of the first name initial and the surname (On et al. 2005).

The algorithms to disambiguate author names that have been proposed up to now differ in several respects (Ferreira, Gonçalves, and Laender 2012). One way to distinguish between different approaches is to classify them as either unsupervised or supervised (Smalheiser and Torvik 2009). Supervised approaches try to train the parameters of a specified model with the help of certain training data (e.g., Torvik and Smalheiser 2009; Ferreira et al. 2010; Levin et al. 2012; Ferreira et al. 2014). The training data contains explicit information as to which author mentions belong to the same identity and which do not. The model trained on the basis of this data is then used to detect relevant patterns in the rest of the data. Unsupervised approaches, on the other hand, try to assess the similarity of author mentions by explicitly specifying a similarity function based on the author mentions' characteristics. We will focus on unsupervised approaches in the following. Supervised approaches entail several problems, especially the challenge of providing adequate, reliable and representative training data (Smalheiser and Torvik 2009).

The unsupervised approaches for disambiguating author names that have been proposed so far vary in several ways. First, every approach specifies a set of attributes and how these are combined to provide a similarity measure between author mentions. In order to determine which similarities are high enough to consider two author mentions or two groups of author mentions as referring to the same author, some form of threshold for the similarity measure is necessary. This threshold can be determined globally for all pairs of author mentions being compared, or it can vary depending on the number of author mentions within a block that refers to a single name representation. Block size dependent thresholds try to reduce the problem of an increasing number of false links for a higher number of comparisons between author mentions, i.e. for larger name blocks (Caron and van Eck 2014; Backes 2018).

Another way in which the approaches differ is the clustering strategy that is applied, i.e. how similar author mentions are grouped together. All clustering strategies used so far in the context of author name disambiguation can be regarded as agglomerative clustering algorithms (Ferreira, Gonçalves, and Laender 2012), especially in the form of single-link or average-link clustering. More specifically, single-link approaches define the similarity of two clusters of author mentions as the maximum similarity of all pairs of author mentions belonging to the different clusters. The idea behind this technique is that each of an author's publications is similar to at least one of his or her other publications. In average-link approaches, on the other hand, the two clusters with the highest overall cohesion are merged in each step, i.e. all objects in the clusters are considered (in contrast to just one from each cluster in single-link approaches). This rests on the assumption that an author's publications form a cohesive entity. As a consequence, it is easier to distinguish between two authors with slightly different oeuvres compared to single-link approaches, but heterogeneous oeuvres by a single author are more likely to be split.

Previous author name disambiguation approaches have usually been evaluated in terms of their quality. This evaluation is always based on measuring how pure the detected clusters are with respect to real-world authors (precision) and how well the author mentions of real-world authors are merged in the detected clusters (recall). However, different metrics have been applied when assessing these properties. Furthermore, different datasets have been used to evaluate author name disambiguation approaches (Kim 2018). It is therefore difficult to compare different approaches based on their previous evaluations.

**Approaches compared**

We focused on unsupervised disambiguation approaches in our analyses (see above). Since these approaches require no training data to be provided a priori, they are more convenient for use with real-world applications. Furthermore, narrowing the set of approaches down to unsupervised ones facilitates their comparison, whereas more aspects have to be considered if they are compared with supervised approaches (e.g., the quality of the training data, which type of supervised model is chosen), making this kind of a comparison more incomprehensible. We chose four approaches in addition to a naïve approach, which only considers the canonical representation of author names. These were selected to cover a wide variety of features that characterize unsupervised approaches for disambiguating author names. We applied the approaches to data from the Web of Science (WoS, Clarivate Analytics) that had already been pre-processed according to a blocking strategy, as described above. More precisely, all author mentions that share the author name representation specified by surname and first initial of the first name have been assigned to the same block. Therefore, all author mentions referring to one real-world author should be in one of these blocks, but there may be several authors represented by one name block. However, there were already some splitting errors in the blocking step (e.g. spelling errors, errors due to name changes).

*Implementation of the four selected disambiguation approaches*

(1) Cota, Gonçalves and Laender (2007) proposed a two-step approach that considers the names of co-authors, publication titles and journal titles. In a first step, all pairs of author mentions that share a co-author name are linked. The linked author mentions are then clustered by finding connected components with regard to this matching. The second step iteratively merges these clusters if they are sufficiently similar with respect to their publication titles or journal titles. The similarity of two clusters (one for publication titles, one for journal titles) is defined as the cosine similarity of the two TF-IDFs (term frequency-inverse document frequency) for the clusters' publication titles (or journal titles). Two clusters are merged if one of their similarities (either with regard to publication titles or to journal titles) exceeds a predefined threshold. This process continues until there are no more sufficiently similar clusters to merge, or until all author mentions are merged into one cluster.

(2) Schulz et al. (2014) proposed a three-step approach based on the following metric for the similarity $s_{ij}$ between two author mentions $i$ and $j$:

$$s_{ij} = \alpha_A \left( \frac{|A_i \cap A_j|}{\min(|A_i|,|A_j|)} \right) + \alpha_S (|p_i \cap R_j| + |p_j \cap R_i|) +$$
$$\alpha_R (|R_i \cap R_j|) + \alpha_C \left( \frac{|C_i \cap C_j|}{\min(|C_i|,|C_j|)} \right) \quad (I)$$

Here, $A_i$ denotes the co-author list of paper $i$, $R_i$ its reference list and $C_i$ its set of citing papers. The first step links all pairs of author mentions with a similarity (determined by formula (I)) exceeding a threshold $\beta_1$ and a set of clusters is determined by finding the corresponding connected components. In the second step, these clusters are merged in a very similar way. In order to determine the similarity $S_{\gamma\kappa}$ of two clusters γ and κ, the similarities between author mentions within these clusters are combined by means of the following formula:

$$S_{\gamma\kappa} = \sum_{i \in \gamma, j \in \kappa} \frac{s_{ij}\Theta(s_{ij})}{|\gamma||\kappa|}, \quad \Theta(s_{ij}) = \begin{cases} 1 \text{ if } s_{ij} > \beta_2 \\ 0 \text{ if } s_{ij} \leq \beta_2 \end{cases} \quad (II)$$

Here, |γ| denotes the number of author mentions in cluster γ (similarly for cluster κ). As the formula shows, only those similarities between author mentions that exceed a threshold $\beta_2$ are considered when calculating the similarity between two clusters. As in the first step, this cluster similarity is used to link clusters if they exceed another threshold $\beta_3$ in order to find the corresponding connected components. The third step of this approach finally adds single author mentions that have not been merged to a cluster in either of the first two steps, provided its similarity with one of the cluster's author mentions exceeds a threshold $\beta_4$.

(3) Caron and van Eck (2014) proposed measuring the similarity between two author mentions based on a set of rules that rely on several paper-level and author-level characteristics. More precisely, a score is specified for each rule, and all of the scores for matching rules are added up to an overall similarity score for the two author mentions (see Table 2). If two author mentions are sufficiently similar with regard to this similarity score, they are linked and the corresponding connected components are considered oeuvres of real-world authors. The threshold for determining whether two author mentions are sufficiently similar depends on the size of the corresponding name block. The idea behind this approach is to take into account the higher risk of false links in larger blocks. Higher thresholds are therefore used for larger blocks to reduce the risk of incorrectly linked author mentions.

Table 2. Rules for rule-based scoring proposed by Caron and van Eck (2014)

| Field | Criterion | Score |
|---|---|---|
| Email | exact match | 100 |
| All initials, more than one | exactly two matching initials | 5 |
| | more than two matching initials | 10 |
| | conflicting initials | -10 |
| First name | matching general name | 3 |
| | matching non-general name | 6 |
| Address (linked to author) | matching country and city | 4 |
| Co-authors | one shared co-author | 4 |
| | two shared co-authors | 7 |
| | more than two shared co-authors | 10 |
| Grant number | at least one shared grant number | 10 |
| Address (linked to publication, but not linked to author) | matching country and city | 2 |
| Subject category | matching subject category | 3 |

| Journal | matching journal | 6 |
|---|---|---|
| Self-citation | at least one publication citing the other | 10 |
| Bibliographic coupling | exactly one shared cited reference | 2 |
| | exactly two shared cited references | 4 |
| | exactly three shared cited references | 6 |
| | exactly four shared cited references | 8 |
| | more than four shared cited references | 10 |
| Co-citation | exactly one shared citing reference | 2 |
| | exactly two shared citing references | 3 |
| | exactly three shared citing references | 4 |
| | exactly four shared citing references | 5 |
| | more than four shared citing references | 6 |

(4) Backes (2018) proposed an approach that starts by considering each author mention as one cluster. An agglomerative clustering algorithm is then employed that iteratively merges clusters if they are sufficiently similar, i.e. two clusters are connected if their similarity exceeds a quality limit $l$. The similarity metric indicating how similar two clusters are takes into account the specificity of the author mentions' metadata. For example, if two author mentions share a very rare subject category this might be a stronger indicator of the author mentions for the same author compared to a very common subject category. This strategy is applied to compute a similarity score for each characteristic under consideration. When using this approach in our study, we considered the following characteristics: titles, abstracts, affiliations, subject categories, keywords, co-author names, author names of cited references, and email addresses. Backes (2018) proposed several variants to combine these scores into a final similarity score of two clusters. In the variant implemented in this study, the scores are combined in the form of a linear combination with equal weights for all characteristics' scores. Each iteration of the clustering process merges all pairs of current clusters whose similarity exceeds $l$. The quality limit $l$ is designed to have a linear dependence on the block size $|author\ mentions|$, whereby the parameter $\lambda$ specifies this relationship (see formula (III)).

$$l = \lambda \cdot |author\ mentions| \qquad (III)$$

*Parameter specification*

Some form of threshold (or a set of thresholds) has to be specified for each of the four approaches. Since such thresholds have not been proposed for all approaches by the authors, and some of the proposed thresholds produce poor results for our dataset, we fitted them with regard to our data. This allows a better comparability since the thresholds are matched to the particular datasets they are applied to. Our procedures for specifying the thresholds maximize the evaluation metrics $F1_{pair}$ and $F1_{best}$ (see below).

We specified such a procedure for each of the approaches that allowed an efficient consideration of a wide range of thresholds. A set of thresholds uniformly distributed over the complete parameter space was chosen as candidate set for the approach of Cota, Gonçalves and Laender (2007). We also specified the thresholds for the approach of Schulz et al. (2014) by evaluating a candidate set of parameters; in this case, the candidate set of thresholds was chosen on the basis of the parameters proposed in the original paper. The parametrization of this approach was further optimized by fitting $\beta_1$, $\beta_2$ and $\beta_3$ independently from $\beta_4$. $\beta_4$ was subsequently chosen based only on the best combination of the other thresholds, which substantially reduces the search space. We believe this to be an adequate procedure for finding the thresholds since the last step of this disambiguation approach (which is based on $\beta_4$) has only a minor influence on the final result. For the approach proposed by Caron and van Eck (2014) we initially had to define the block size classes that divide the blocks into several classes with regard to the number of author mentions in them. Similar to Caron and van Eck (2014), we defined six block size classes. Our specification of the classes aims at reducing the variance of optimal thresholds within a class.

For the approach of Backes (2018), we had to modify the approach slightly in order to define a feasible procedure for fitting the parameter $\lambda$, which determines the quality limit $l$ for a given block. Instead of linking all pairs of clusters whose similarity exceeds a given $l$ in each iteration, we iteratively merged only those pairs of clusters whose similarity equals the maximum similarity of all current pairs of clusters (the clusters are recomputed after each merger). These similarities were taken as estimates for the quality limit that would yield the clustering of the corresponding merger step. This modification may produce results that are different to the original approach, since the order in which the author mentions are merged may change and the similarities between clusters depend on the previous mergers. However, we assume that these changes would produce only minor differences that do not influence any general conclusions on the approach. Our implementation merges the most similar clusters in each iteration, i.e. the most reliable mergers are applied iteratively until the quality limit is reached. Correspondingly, the original approach follows the idea that all cluster similarities exceeding a certain quality limit indicate reliable links between the corresponding clusters.

**Data**

We collected metadata for a subset of author mentions from the WoS for our analyses. In order to provide a gold standard that represents sets of author mentions corresponding to real-world authors, we only took author mentions with a ResearcherID linked to them in the WoS into account. More specifically, all person records that are marked as authors and that have a ResearcherID linked to at least one paper published in 2015 or later have been considered. It is very likely that this procedure excludes all author mentions with ResearcherIDs referring to non-author entities (e.g. organizations) and takes into account only such ResearcherIDs that have been maintained recently. We applied the same standardization for all name-based metadata as was used to block author mentions, i.e. a canonical name representation is used consisting of first name initial and surname. We only considered name blocks comprising at least five real-world authors. This selection allowed us to focus on rather difficult cases where the author mentions in a block actually have to be disambiguated across several authors. All in all, this data collection procedure results in 1,057,978 author mentions distributed over 2,484 name blocks and 29,244 distinct ResearcherIDs. The largest name block ("y. wang") comprises 7,296 author mentions.

**Results**

*Evaluation metrics*

The evaluation of author name disambiguation approaches is generally based on assessing their ability to discriminate between the author mentions of different real-world authors (precision) and their ability to merge the author mentions of one real-world author (recall). Even though these concepts are widely accepted and referenced, different specific evaluation metrics have been used in the past. In the following, we focus on two types of evaluation metrics. First, we calculate the pairwise precision ($P_{pair}$), pairwise recall ($R_{pair}$) and pairwise F1 ($F1_{pair}$) (Levin et al. 2012; Caron and van Eck 2014; Backes 2018) for each of the approaches. Whereas the pairwise precision measures how many of the links between author mentions in the detected clusters are correct, the pairwise recall measures how many of the links between author mentions of real-world authors are correctly detected. Pairwise F1 is the harmonic mean of these two metrics. In formulae (IV)-(VI), $pairs_{author}$ denotes the set of pairs of author mentions where both of the author mentions refer to the same author, and $pairs_{cluster}$ denotes the set of pairs of author mentions where both author mentions are assigned in the same cluster by the disambiguation algorithm. Each of the pairwise evaluation metrics can take values between 0 (no true links between author mentions detected) and 1 (all true links between author mentions detected).

$$P_{pair} = \frac{|pairs_{author} \cap pairs_{cluster}|}{|pairs_{cluster}|} \quad \text{(IV)}$$

$$R_{pair} = \frac{|pairs_{author} \cap pairs_{cluster}|}{|pairs_{author}|} \quad \text{(V)}$$

$$F1_{pair} = \frac{2P_{pair}R_{pair}}{P_{pair}+ R_{pair}} \quad \text{(VI)}$$

Second, we calculate metrics to measure how reliably a cluster can be attributed to one specific author (best precision $P_{best}$) and how well an author can be attributed to one specific cluster (best recall $R_{best}$).

More specifically, the best precision represents the fraction of author mentions that refer to the most represented author in the corresponding cluster. The most represented author of a cluster is defined as the author with the largest group of author mentions in this cluster. Accordingly, the best recall represents the fraction of author mentions that are assigned to the cluster with the most author mentions of the corresponding author. Similar to the pairwise F1, the best F1 $F1_{best}$ combines best precision and best recall in the form of their harmonic mean. In formulae (VII)-(IX), $author\ mentions_{best\ author}$ denotes the set of author mentions referring to the author most of the corresponding cluster's author mentions refer to, $author\ mentions_{best\ cluster}$ denotes the set of author mentions assigned to the cluster with the most author mentions of the corresponding author and $author\ mentions$ denotes the set of all author mentions. Technically speaking, $P_{best}$, $R_{best}$ and $F1_{best}$ can also take values between 0 and 1. However, $author\ mentions_{best\ author}$ and $author\ mentions_{best\ cluster}$ will always contain at least one author mention. Actually, these evaluation metrics will thus always be greater than 0.

$$P_{best} = \frac{|author\ mentions_{best\ author}|}{|author\ mentions|} \quad \text{(VII)}$$

$$R_{best} = \frac{|author\ mentions_{best\ cluster}|}{|author\ mentions|} \quad \text{(VIII)}$$

$$F1_{best} = \frac{2P_{best}R_{best}}{P_{best}+ R_{best}} \quad \text{(IX)}$$

Each of these formulae can either be applied to the complete dataset or to a subset of author mentions. For example, the results of one name block can be evaluated by only considering author mentions within this block when computing the evaluation metrics.

*Overall results*

The results for the approaches described above are summarized in Table 3. The table shows the evaluation metrics described in the previous section for all of the approaches. All of the approaches produce better results than the naïve baseline disambiguation. The approach proposed by Caron and van Eck (2014) performs best among the examined approaches with regard to both $F1_{pair}$ and $F1_{best}$. If one compares the approaches of Schulz et al. (2014) and

Backes (2018), the two evaluation metrics yield different rankings. Whereas the latter approach performs better with regard to $F1_{pair}$, the first performs better with regard to $F1_{best}$. Finally, the approach of Cota, Gonçalves, and Laender (2007) performs only slightly better than the baseline disambiguation. The precision in particular is very low in this case, due mainly to a high number of false links between author mentions in the first step (merging author mentions with shared co-authors).

**Table 3. Overall results for all approaches**

| Approach | $P_{pair}$ | $R_{pair}$ | $F1_{pair}$ | $P_{best}$ | $R_{best}$ | $F1_{best}$ |
|---|---|---|---|---|---|---|
| Baseline | 0.095 | 1.000 | 0.173 | 0.322 | 1.000 | 0.487 |
| Cota, Gonçalves, and Laender (2007) | 0.111 | 0.858 | 0.196 | 0.442 | 0.913 | 0.596 |
| Schulz et al. (2014) | 0.453 | 0.457 | 0.455 | 0.799 | 0.750 | 0.773 |
| Caron and van Eck (2014) | 0.831 | 0.787 | 0.808 | 0.916 | 0.885 | 0.900 |
| Backes (2018) | 0.674 | 0.622 | 0.647 | 0.761 | 0.699 | 0.729 |

Figure 1 shows the distribution of the disambiguation quality over block sizes (the mean of all blocks of a specific size is plotted). This distribution is shown for the case where the thresholds are specified as described above ("original") and for the case where the optimal thresholds for each single block are used ("flexible"). The results reveal that the disambiguation quality varies strongly across name blocks. The quality generally worsens for large blocks. Therefore, the disambiguation process may produce biases with regard to the frequency of the corresponding name representation. One reason for the disambiguation quality's dependence on the size of the name block is the larger search space to find clusters of author mentions. This increases the search complexity in general, implying a greater potential for false links between author mentions. Some approaches try to reduce this problem by allowing for block size dependent thresholds (see next section). Even though the negative relationship between block size and disambiguation quality can be observed for all approaches, the decline in quality is not equal in all of them. Especially for the approach of Caron and van Eck (2014), the influence of the block size is relatively small.

*Influence of parametrization on the disambiguation quality*

Among the approaches included in our comparison, Caron and van Eck (2014) and Backes (2018) used block size dependent thresholds. As described above, the first approach is based on defining one threshold for each of six block size classes, whereas the threshold is linearly dependent on the block size in the second approach. Table 4 shows the block size classes and corresponding thresholds used by our implementation for the approach of Caron and van Eck (2014). In contrast, the approaches of both Cota, Gonçalves, and Laender (2007) and Schulz et al. (2014) use global thresholds for all block sizes.

**Table 4. Block size classes and thresholds for Caron and van Eck (2014)**

| Block size | Threshold ($F1_{pair}$) | Threshold ($F1_{best}$) |
|---|---|---|
| 1-500 | 21 | 19 |
| 501-1000 | 22 | 21 |
| 1001-2000 | 25 | 23 |
| 2001-3000 | 27 | 25 |
| 3001-4500 | 29 | 25 |
| >4500 | 29 | 27 |

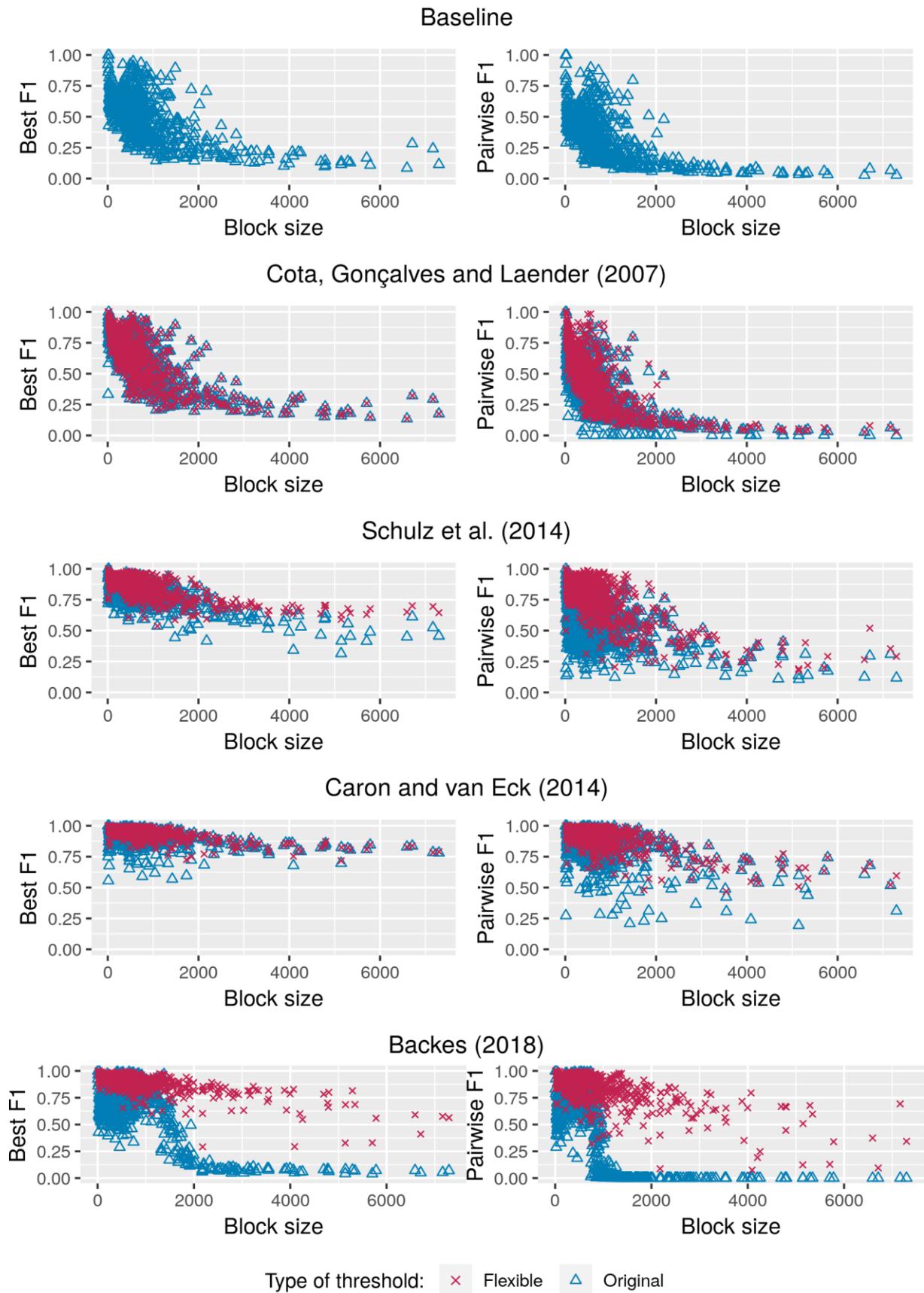

**Figure 1. Distribution of disambiguation quality over block sizes**

In Figure 1, the results based on optimal thresholds for each single block (flexible thresholds) represent an upper bound for the quality over all possible thresholds. Flexible thresholds would not greatly improve the quality of the approach of Cota, Gonçalves, and Laender (2007) since the results based on global thresholds are very close to the results based on completely flexible thresholds. The reason for this is that the quality is dominated by the first step of the approach, which does not employ any threshold at all. The second step, on the other hand, does not change the results significantly, so that the effect of the thresholds is rather small. In contrast, the approach of Schulz et al. (2014) could benefit from using flexible thresholds, especially for large blocks.

Similar to the approach of Cota, Gonçalves, and Laender (2007), the difference between the original implementation and the one with flexible thresholds is rather small for the approach of Caron and van Eck (2014). However, the choice of thresholds does affect the result in this case, as shown by the comparison with an implementation based on a constant threshold for all block sizes. Table 5 shows the evaluation results for the approach of Caron and van Eck (2014) with three different types of thresholds: a constant threshold for all blocks ("Constant"), the thresholds of the block size classes shown in Table 4 ("Block size classes"), and the optimal threshold for each single block ("Flexible"). These results show that the original implementation produces better results than those obtained using a constant threshold. This means that the somewhat rough partitioning between six block size classes already allows for an adequate differentiation with regard to the threshold, and that this strategy improves the disambiguation result compared to a constant threshold over all block sizes. In contrast, the strategy of specifying a threshold which is linearly dependent on the block size, as employed by the approach of Backes (2018), is unable to find good thresholds over the complete range of block sizes. This is due mainly to a drop in the recall (together with an increasing precision) for large blocks. The thresholds chosen by the algorithm are thus too high for large blocks. Hence, a linear relationship between block size and threshold would not appear to be an adequate strategy for large blocks. The fitted thresholds for the approach of Caron and van Eck (2014) also confirm that a nonlinear relationship between block size and threshold may be more suitable.

Table 5. Results for different types of thresholds for Caron and van Eck (2014)

| Type of threshold | $P_{pair}$ | $R_{pair}$ | $F1_{pair}$ | $P_{best}$ | $R_{best}$ | $F1_{best}$ |
|---|---|---|---|---|---|---|
| Constant | 0.690 | 0.741 | 0.714 | 0.879 | 0.880 | 0.880 |
| Block size classes | 0.831 | 0.787 | 0.808 | 0.916 | 0.885 | 0.900 |
| Flexible | 0.907 | 0.850 | 0.878 | 0.954 | 0.897 | 0.924 |

The results in Figure 1 and Table 5 demonstrate that the disambiguation quality can be improved if flexible thresholds dependent on the block size are specified. However, the specification of adequate thresholds is generally a non-trivial task since it depends on the data at hand. Likewise, the thresholds proposed previously for the approaches examined in this paper do not correspond to the thresholds fitted with regard to our dataset.

**Discussion**

The disambiguation of units (researchers, research groups, institutions etc.) for bibliometric analyses is an important topic in research evaluation. The results of evaluation studies can only be as good as the underlying data. For example, Clarivate Analytics annually publishes the names of highly cited researchers who have published the most papers belonging to the 1% most highly cited in their subject categories (see https://hcr.clarivate.com). The reliable attribution of papers to corresponding researchers is an absolute necessity for publishing this list of researchers. Although different disambiguation approaches have been developed and implemented in local bibliometric databases (e.g., Caron and van Eck 2014), there is hardly any comparison of the approaches. However, this comparison is necessary to obtain indicators of the best approaches, or those conditions on which the performance of the approaches depends. In this paper, we compared different author name disambiguation approaches based on a dataset containing author identifiers in the form of ResearcherIDs. This allows a better comparison of different approaches than if previous evaluations are used since these are generally based on different databases. Our results show that all of the approaches included in the comparison perform better than a baseline that only uses a canonical name representation of the authors for disambiguation. Although the comparison does not point to the recommendation of one approach for all disambiguation tasks, it does provide evidence of when which approach can produce good results – especially with regard to the size of corresponding name block sizes. As our analyses show, the parametrization of the approaches can have a significant effect, which depends largely on the data at hand. Therefore, the context of the disambiguation task has to be taken into account for a proper implementation of an algorithm. In the context of this study, the approach proposed by Caron and van Eck (2014) produced the best results.

Future research should further examine how different author name disambiguation approaches behave and how certain features affect the disambiguation results. For example, the set of characteristics used by the approaches may play an important role. Since the approaches included in our comparison use different sets of characteristics, differences in the results may be due in part to the choice of the characteristics used. A more detailed analysis of this choice in future studies may shed more light on which set of characteristics is most suitable for which context.

Understanding how author name disambiguation approaches behave is important in order to improve the algorithms and to assess the effect they have on analyses building on the disambiguated data. A good understanding of the behaviour is the basis for reliable analyses at the individual level.

## Acknowledgments

The bibliometric data used in this paper are from an in-house database developed and maintained in cooperation with the Max Planck Digital Library (MPDL, Munich) and derived from the Science Citation Index Expanded (SCI-E), Social Sciences Citation Index (SSCI), Arts and Humanities Citation Index (AHCI) prepared by Clarivate Analytics, formerly the IP & Science business of Thomson Reuters (Philadelphia, Pennsylvania, USA). We would like to thank Robin Haunschild and Thomas Scheidsteger from the Central Information Service for the institutes of the Chemical Physical Technical (CPT) Section of the Max Planck Society (IVS-CPT) for providing the computational infrastructure for conducting our analyses.